\documentclass[preprint,nofootinbib]{revtex4-1}

\usepackage{amsmath}
\usepackage{amsfonts,amssymb}
\usepackage{hyperref}
\usepackage{graphicx}
\usepackage[utf8]{inputenc}

\begin{document}

\title{Late-times asymptotic equation of state for a class of nonlocal theories of gravity}

\author{Leonardo Giani$^{1,2,3}$}
\email{giani@thphys.uni-heidelberg.de}

\author{Oliver F. Piattella$^{1,2,3}$}
\affiliation{$^1$ Department of Physics, Universidade Federal do Esp\'irito Santo, Avenida Fernando Ferrari 514, 29075-910 Vit\'oria, Esp\'irito Santo, Brazil,}

\affiliation{$^2$ N\'ucleo Cosmo-ufes and PPGCosmo, Universidade Federal do Esp\'irito Santo, Avenida Fernando Ferrari 514, 29075-910 Vit\'oria, Esp\'irito Santo, Brazil,}

\affiliation{$^3$ Institut f\"ur Theoretische Physik, Ruprecht-Karls-Universit\"at Heidelberg, Philosophenweg 16, 69120 Heidelberg, Germany}

\begin{abstract}
 
We investigate the behavior of the asymptotic late-times effective equation of state for a class of nonlocal theories of gravity. These theories modify the Einstein-Hilbert Lagrangian introducing terms containing negative powers of the d'Alembert operator acting on the Ricci scalar. We find that imposing vanishing initial conditions for the nonlocal content during the radiation-dominated epoch implies the same asymptotic late-times behavior for most of these models. In terms of the effective equation of state of the universe, we find that asymptotically $\omega_{ \rm eff} \rightarrow -1$, approaching the value given by a cosmological constant. On the other hand, unlike in the case of $\Lambda$CDM, the Hubble factor is a monotonic growing  function that diverges asymptotically.  We argue that this behavior is not a coincidence and discuss under which conditions this is to be expected in these nonlocal models.
\end{abstract}

\maketitle

\section{Introduction}

The sundown of the last century greeted physicists with the discovery of the accelerated expansion of the universe \cite{Riess:1998cb}\cite{Perlmutter:1998np}, setting up what turned out to be one of the most challenging puzzles of the subsequent 20 years. From a pure phenomenological point of view the puzzle is simply resolved by introducing in the universe a new source of energy-momentum in the right hand side of the Einstein field equations. Another way around is to consider a cosmological constant $\Lambda$ in their left hand side, and attribute the accelerated expansion of the universe to a geometrical effect. In both cases, for them to be compatible with the observed acceleration of the universe, such components generally dubbed Dark Energy (DE) should represent almost the 70 percent of the total density budget. The overall picture looks even worse if we consider that in order to explain consistently structure formation our best bet is to invoke the existence of another cold and weakly interacting matter component, dubbed Cold Dark Matter (CDM), which accounts for another 25 percent of the density of the universe \cite{Ade:2015xua}\cite{Aghanim:2018eyx}; the resulting model is called concordance model, or standard model of cosmology, and dubbed $\Lambda$CDM. This paradigm, however, is based on our faith that General Relativity (GR) is the correct description of the gravitational interaction; relaxing this assumption can lead to different scenarios \cite{1983:Mond}, even if the the price to pay is maybe too high being it giving up from one of the most experimentally successful and beautiful theories of the last century. 

In this work we are interested in models that avoid the introduction of a cosmological constant and try to explain the presence of an effective DE by introducing modifications of the standard Einstein-Hilbert Lagrangian. Such attempts are motivated by the fact that a cosmological constant $\Lambda$, even being a very simple and effective source for the accelerated expansion of the universe, is not completely satisfactory from the theoretical point of view. For an overview of the conceptual issues related to the cosmological constant see Refs. \cite{RevModPhys.61.1}, \cite{MARTIN2012566}.

If we try to put DE as a matter source in the Einstein field equations, e.g. we put it in the right hand side as a contribution to the stress-energy tensor, we are assuming the existence of a very strange cosmological fluid which exhibits an exotic behavior, it possesses negative pressure. To overcome this problem, DE is usually addressed by considering modifications of GR that act on the geometrical sector, i.e. in the left hand side of the Einstein field equations. A very interesting class of these theories are generally dubbed as $f(R)$ theories; in such kind of models instead of the Einstein-Hilbert Lagrangian in the action functional we have some arbitrary function of the Ricci scalar, which then is usually constrained by requiring that  GR is recovered at scales in which we trust it, e.g. solar system scales. For a review on the topic see for example Refs.  \cite{DeFelice2010,Capozziello:2009nq, CAPOZZIELLO2011167}.
With the idea of going beyond GR without spoiling its success a number of attempts were made by relaxing some of its basic principles; some of the most popular ones include abandoning Lorentz invariance, the equivalence principle and the general covariance principle. Some examples are Ho\u rava-Lifshitz gravity \cite{Horava:2009uw}, unimodular gravity and its extensions \cite{VANDERBIJ1982307,HENNEAUX1989195,Barvinsky:2017pmm}. For a general review on the topic see for example  Ref. \cite{Nojiri:2010wj}

An interesting class of $f(R)$ theories that has become popular in the last years prescribes the modification of the action introducing terms of nonlocal nature like the inverse of the d'Alembert operator acting on the Ricci scalar, see for example Ref. \cite{Jhingan:2008ym}. Some of these theories are for example the Deser Woodard (DW) model \cite{Deser:2007jk} and the $RR$ model proposed by Maggiore and Mancarella  \cite{Maggiore:2014sia}. Other interesting nonlocal theories are defined instead at the level of the field equations and still lack a Lagrangian formulation, a prototypical example being the $RT$ model \cite{Maggiore:2013mea}. 

In Ref. \cite{Nersisyan:2016hjh} the authors perform a dynamical system analysis of the $RR$ model; numerical investigations show in particular that independently from the value of the only free parameter of the theory (when the initial conditions on the local fields are compatible with the standard cosmological history up to the epoch of matter domination), the effective equation of state of the universe approaches at late-times the value $\omega_{\rm eff} \rightarrow -1$. Later on, in Ref. \cite{Vardanyan:2017kal}, another nonlocal model of this fashion is proposed and investigated; it turns out that in its simplest version the model contains only one free parameter, and numerical investigations show that independently from its specific value the asymptotic effective equation of state once again approaches the value $\omega_{\rm eff} \rightarrow -1$. The authors of Ref.  \cite{Vardanyan:2017kal} remark this behavior and point out its similarity with the $RR$ model. Later on, in Ref. \cite{Giani:2019vjf}, a dynamical system analysis of the model is performed and it is proved analytically that imposing initial conditions compatible with a radiation-dominated epoch determines asymptotically that $\omega_{\rm eff} \rightarrow -1$.

In this work we apply the technique used in Ref. \cite{Giani:2019vjf} to other nonlocal theories and study the behavior of their late-times asymptotic equations of state. We stress that our results are analytical, and confirm the numerical investigations already present in the literature for specific models. 

The structure of the paper is the following: in Sec. \ref{section2} we briefly introduce the nonlocal models we are working with and their cosmological field equations; in Sec. \ref{section3} we compute their  asymptotic equation of state. Finally, in Sec. \ref{section4} we present our conclusions and discuss our results.

\section{The nonlocal models}\label{section2}
In this section we introduce the nonlocal models for which we study the asymptotic behavior of the effective equation of state. They share the property of being defined in terms of functions of the inverse of the d'Alembertian operator acting on the Ricci scalar which are added to the Einstein Hilbert Lagrangian.\footnote{These are the only kind of nonlocal models which we treat in this work; we leave for future investigation  similar models built with the use of nonlocal operators acting on the Ricci tensor, like the one proposed in Ref. \cite{Barvinsky:2011hd}, or that are obtained by modifying the field equations instead of the action functional, like the $RT$ model.}
Such kind of nonlocal terms which are normally absent in classical theories are instead generally present in the quantum effective action, so they can manifest themselves in a consistent theory of quantum gravity. For a general review on the topic see for example \cite{Vilkovisky:1992pb}; for a specific treatment for the models of interest in this work we address the reader to Ref. \cite{Belgacem:2017cqo}.

It is in general possible to rewrite such nonlocal theories in a localized form by introducing a number of auxiliary fields. This technique was created and applied the first time for the DW model in Ref. \cite{Nojiri:2007uq}, however its extension to other nonlocal theories is straightforward and was used in almost all the subsequent works involving similar nonlocal theories as a helpful computational tool. It is important to stress that such parametrization must be carefully achieved with particular attention to the initial conditions on the localized fields, since if the latter are not properly given then spurious propagating degrees of freedom might appear.

In this work we address the late-times cosmological implications of a list of nonlocal modifications of the Einstein-Hilbert action; it turns out that it is more convenient to work within the framework of localized fields.

We assume a FLRW spatially flat background metric so that the line element is given by:
\begin{equation}
    ds^2 = -dt^2 + a^2(t)\delta_{ij}dx^idx^j \; ,
\end{equation}
where, $a(t)$ is the scale factor. We work using the e-fold time parameter $N \equiv \log{ a}$ and the  Hubble factor normalized to the Hubble constant $h \equiv H/H_0$; the matter and radiation densities are defined as:
\begin{equation}
   \Omega_R \equiv \frac{8\pi G\rho_R}{3H_0^2}\equiv \Omega^{0}_{R}e^{-4N} \; , \qquad \Omega_M\equiv \frac{8\pi G\rho_M}{3H_0^2}\equiv \Omega^{0}_{M}e^{-3N} \; ,
\end{equation}
where $\Omega^0_M$ and $\Omega^0_R$ are the values of the matter and radiation densities today.\footnote{As it is customary the scale factor is normalized in such a way that today, $t=t_0$, $a(t_0) = 1$ so that $N_0=0$.}

We also recall the definition of the effective equation of state of the universe:
\begin{equation}
    \omega_{\rm eff} = -1 -\frac{2}{3} \xi \; ,
\end{equation} 
 where $\xi \equiv h'/h$, with the prime denoting derivation with respect to $N$, parametrizes its time evolution.
 
In the next subsections we briefly present the models we are dealing with and their cosmological background equations, together with the definitions of the auxiliary fields and their Klein-Gordon (KG) equations.

To make easier the comparison between the equations of different models we use the same letters $U,V,W,Z$ to label the localized fields, in particular $U$ is always defined by $\Box U = - R$. 
All the models considered are characterized by a single coupling parameter, and in the background equations we employ the same symbol $\gamma$ to define powers of these parameters in units of the Hubble factor today.

Finally, for each model, we briefly discuss how the request of a radiation-dominated epoch affect the choice of initial conditions for the auxiliary fields.

\subsection{The $RR$ model}
To begin with let us introduce the $RR$ model, proposed by Maggiore and Mancarella in Ref. \cite{Maggiore:2014sia}. For a general review on the model we address the reader to Ref.  \cite{Belgacem:2017cqo}; the cosmological perturbation theory and the impact on structure formation are studied in Ref. \cite{Dirian:2014ara}. A dynamical system analysis of the model is numerically performed in Ref.  \cite{Nersisyan:2016hjh}, while in  Ref. \cite{Dirian:2016puz} the model is tested against observation and compared with the $\Lambda$CDM.

In this theory one adds to the usual Einstein-Hilbert Lagrangian a nonlocal term of the form:
\begin{equation} 
    \mathcal{L}= \mathcal{L}_{EH} -\frac{1}{6}m^2 R\frac{1}{\Box^2}R \;.
    \end{equation}
Introducing the auxiliary fields 
\begin{eqnarray}
        \Box U = -R \; , \\
        \Box S = -U \; , 
\end{eqnarray}
defining the dimensionless quantity $V = H_0^2S$ and by varying the action with respect to the metric tensor we obtain the following cosmological equations \cite{Nersisyan:2016hjh}:
\begin{eqnarray}
        h^2 = \frac{\Omega_{M}^0e^{-3N} + \Omega_{R}^0e^{-4N} +\frac{\gamma}{4}U^2}{1 + \gamma\left(-3V -3V' + \frac{1}{2}U'V'\right)} \; , \label{CosmoRR1} \\
        \xi = \frac{\frac{-3\Omega_M -4\Omega_R}{h^2} + 3\gamma\left(\frac{U}{h^2} + U'V' -4V'\right)}{2\left(1-3\gamma V\right)} \label{CosmoRR2} \; ,
\end{eqnarray}
where we have defined $\gamma \equiv m^2/9H_0^2$. The KG equations obtained varying the action with respect to the auxiliary fields are:
\begin{eqnarray}
        V'' + V'\left(3 + \xi\right)= \frac{U}{h^2}\; , \label{KGRRV} \\
        U'' + U'\left(3+\xi \right)= 6\left(2 + \xi \right) \label{KGRRU} \; .
\end{eqnarray}

If we impose initial conditions compatible with a radiation-dominated epoch, i.e. $h_i^2 \sim \Omega_{Ri}$ and $\xi_i \sim -2$, Eqs. \eqref{CosmoRR1} and \eqref{CosmoRR2} at some initial time $N = N_i$ become:
\begin{eqnarray}
\frac{U_i^2}{4h_i^2}&=&  -3V_i -3V_i' + \frac{1}{2}U_i'V_i' \; ,\label{ICCosmoRR1}\\
V_i &=& \frac{U_i}{4h_i^2} + \frac{1}{4}U_i'V_i' - V_i' \;  \label{ICCosmoRR2},
\end{eqnarray}
so that they provide two constraints for the four initial conditions required on $V_i, V_i', U_i, U_i'$. We choose vanishing initial conditions for all of them, thereby satisfying in natural way the constraints of Eqs. \eqref{ICCosmoRR1} and \eqref{ICCosmoRR2}.

\subsection{The $\Box^{-1}R$ model}
This model was proposed in Ref. \cite{Vardanyan:2017kal}, where the possibility of introducing nonlocal operators in a bimetric theory of gravity was investigated for the first time. However, it turns out that in the simplest version of the theory the bimetric nature of the model is not explicit and the action functional reduces to the standard Einstein-Hilbert term plus a nonlocal deformation of the form:
\begin{equation}
    \mathcal{L} = \mathcal{L}_{EH} + m^2\frac{1}{\Box}R \; .
\end{equation}
Later on, in Ref. \cite{Amendola:2017qge}, this model was introduced in a class of models motivated from studies of nonperturbative lattice quantum gravity. Its background cosmology is numerically studied in Ref. \cite{Vardanyan:2017kal} and here it is showed its compatibility with the cosmological history of the $\Lambda$CDM. A dynamical system analysis of the model and its Newtonian limit are performed in Ref. \cite{Giani:2019vjf}.
Introducing the auxiliary fields:\footnote{Differently from Refs. \cite{Vardanyan:2017kal, Giani:2019vjf} we choose to define $U$ with a minus sign in such a way that all the model considered in this work have the same KG equation for $U.$}
\begin{eqnarray}
        \Box U = -R \; , \\
        \Box V = -m^2 \; ,
\end{eqnarray}
 defining $\tilde V \equiv 1-V$ and $\gamma \equiv m^2/H_0^2$ the KG equations for the auxiliary fields and the background cosmological equations become \cite{Giani:2019vjf}:
\begin{eqnarray}
3\tilde V - \frac{\gamma U}{2h^2} + 3\tilde V' - \frac{U'\tilde V'}{2} = \frac{3\Omega_{R} + 3\Omega_M}{h^2}\;, \label{CosmoVAAS1}\\
\label{acceq} -\tilde V\left(3 + 2\xi\right) + \frac{\gamma}{h^2}(1 + U/2) + \tilde V' - \frac{U'\tilde V'}{2} = \frac{\Omega_{R}}{h^2}\;, \label{CosmoVAAS2}\\
U'' + \left(3 + \xi\right)U' - 6\left(2 + \xi\right) = 0\;,\label{KGVAASU}\\
\tilde V'' + \left(3 + \xi\right)\tilde V' = -\frac{\gamma}{h^2}\; \label{KGVAASV}.
\end{eqnarray}
Imposing initial conditions compatible with a radiation-dominated era, i.e. $h_i^2 \sim \Omega_{Ri}$ and $\xi_i \sim -2$ Eqs. \eqref{CosmoVAAS1} and \eqref{CosmoVAAS2} at some initial time $N=N_i$ read:
\begin{eqnarray}
        \tilde{V_i} - \frac{\gamma U_i}{6h^2} +\tilde V_i' -\frac{1}{6}U_i'\tilde V_i' \; , \label{ICCosmoVAAS1} \\
        \tilde V_i + \frac{\gamma}{h_i^2}\left(1 + \frac{U_i}{2}\right) + \tilde V_i' -\frac{1}{2}U_i'\tilde V_i' = 1 \; . \label{ICCosmoVAAS2}
\end{eqnarray}
The latter equations provide two constrains among the four initial conditions on the auxiliary fields $U_i,U_i',\tilde{V}_i,\tilde{V}_i'$. We make the natural choice $U_i = U_i'= \tilde{V}_i' = 0$ and $\tilde{V}_i = 1$. Note that in order to satisfy the above constraints we need to assume $\gamma^2/h_i^2 \ll 1$, which is however reasonable since $h_i^2 \sim e^{-4N_i}$ and $N_i$ is very large and negative.

\subsection{The $\Box^{-2}R$ model}
Like the previous one, also this model is motivated from studies of nonperturbative lattice quantum gravity. It was proposed in Ref.  \cite{Amendola:2017qge} where its background cosmology is studied. It is obtained by adding to the Einstein-Hilbert action a nonlocal term of the form:
\begin{equation}
      \mathcal{L} = \mathcal{L}_{EH} -\frac{M^4}{6}\frac{1}{\Box^2}R \; .
\end{equation}
In order to localize the theory it is necessary to define four auxiliary fields:
\begin{eqnarray}
\Box U = -R\; , \\
\Box S = -U \; , \\
\Box Q = -1\; ,  \\
\Box L = -Q \; .
\end{eqnarray}
Introducing the parameter $\gamma \equiv M^4/9H_0^4$ and defining $V \equiv H_0^2 S , W\equiv H_{0}^2Q, Z \equiv H_0^4L$  the cosmological and the KG field equations become:
\begin{eqnarray}
        h^2=\frac{\gamma}{4}\left[V + WU + h^2\left( 6Z + 6Z' -U'Z' -V'W' \right)\right] + \Omega_R^0e^{-4N} + \Omega_M^0 e^{-3N} \; , \label{CosmoB1} \\
        \xi = \frac{1}{2\left(1 - \frac{3}{2}\gamma Z\right)}\left[\frac{-4\Omega_R^0e^{-4N}  -3\Omega_M^0 e^{-3N}}{h^2} + \frac{3}{2}\gamma \left(\frac{W}{h^2} -4Z' + U'Z' + V'W'\right)\right] \; , \label{Cosmob2}
        \end{eqnarray}
        \begin{eqnarray}
        U'' + \left(3 + \xi\right)U = 6\left(2 + \xi\right) \;, \label{KGBU} \\
        V'' + \left(3 + \xi\right)V'= \frac{U}{h^2}\;, \label{KGBV}\\
        W'' + \left(3 + \xi\right)W'= \frac{1}{h^2}\;, \label{KGBW}\\
        Z'' + \left(3 + \xi\right)Z'= \frac{W}{h^2}\;   \label{KGBZ}.
\end{eqnarray}
If we impose initial conditions compatible with a radiation-dominated epoch, in which $\xi_i \sim -2$ and $h_i^2 \sim \Omega_{Ri}$, Eqs. \eqref{CosmoB1} and \eqref{Cosmob2} read:
\begin{eqnarray}
        h^2\left(6Z_i + 6Z_i' -U_i'Z_i' + V_i'W_i' \right) = -V_i - W_iU_i \; , \label{ICCosmoB1}\\
        -4 = \frac{1}{1 - \frac{3}{2}\gamma Z_i}\left[1 + \frac{3}{2}\gamma\left(\frac{W_i}{h^2} - 4Z_i' + U_i'Z_i' + V_i'W_i'\right) \right] \; \label{ICCosmoB2},
\end{eqnarray}
and we are left with two constraints for eight initial conditions on the auxiliary fields. We make the natural choice $U_i = V_i = Z_i = U_i' = V_i' = Z_i ' = 0$ to satisfy Eq. \eqref{ICCosmoB1}. Inserting these initial conditions in Eq. \eqref{ICCosmoB2} we are left with: 
\begin{equation}
    \frac{1}{2}\left(-4 + \frac{3}{2}\gamma\frac{W_i}{h_i^2} \right) = -2 \; .
\end{equation}
The latter is satisfied for $W_i = 0$, however we can choose other values of $W_i$ as long as $|W_i| \ll h_i^2$, while we have no constraint at all on $W_i'$.  We choose to set the initial conditions $W_i = W_i' = 0$, but the qualitative analysis of the model at late-times of the following section is not affected if we choose any other positive value for $W_i, W_i'$.

\subsection{The Deser-Woodard model}
This model was proposed in Ref. \cite{Deser:2007jk} and represents an attempt to incorporate nonlocal gravitational effects without assuming a priori a specific form for the nonlocal operator. The idea is to introduce a free function, called distortion function, of the $\frac{1}{\Box}R$ operator in the action, to constrain such function in order produce a cosmological history identical to the one of the $\Lambda$CDM (but with no cosmological constant), and finally, once that the background and the free function are chosen, look for testable predictions.
For a review on the main features of the model we address the reader to Ref. \cite{Woodard:2014iga}; the issue of ghosts is studied in Ref. \cite{Park:2019btx}.
A detailed study of its dynamics is performed in Ref. \cite{Koivisto:2008xfa}, while its  Newtonian limit is studied in Ref. \cite{Koivisto:2008dh}. The effects of such kind of modification for structure formation are studied in Refs. \cite{Park:2012cp,Dodelson:2013sma,Nersisyan:2017mgj}, while constraints from observational datasets are found in Ref. \cite{Amendola:2019fhc}. Finally, an improved version of the model has been recently proposed in Ref. \cite{Deser:2019lmm}.

The Lagrangian density is given by:
\begin{equation}
    \mathcal{L} = \mathcal{L}_{EH} + Rf\left(\frac{1}{\Box}R\right) \; ,
\end{equation}
and its localized form was obtained for the first time in Ref. \cite{Nojiri:2007uq} by defining the auxiliary fields:\footnote{We are using a different definitions for the field $U$ with respect to the one of Ref. \cite{Nojiri:2007uq} in order have the same KG equation for the field $U$ in all the models considered in this work. They are obtained by making the substitutions $U \rightarrow -U$, $\Bar{f}\rightarrow-\Bar{f}$. }
\begin{eqnarray}
        \Box U = -R \; ,\\
        \Box V = \Bar{f}(U)R \; ,
\end{eqnarray}
where the symbol $\Bar{f}$ is used to indicate the derivative of the distortion function $f$ with respect to $U$. The cosmological background equations are:
\begin{eqnarray}
        \left(1 + f - V\right) = -\frac{U'V'}{6} - f' + V' +\frac{\Omega_R + \Omega_M}{h^2} \; , \label{CosmoDW1}\\
        \left(2\xi + 3\right)\left(1 + f - V\right) = V'' - f'' + \left(V' - f'\right)\left(2 + \xi\right) + \frac{U'V'}{2} - \frac{\Omega_R}{h^2} \; ,\label{CosmoDW2}
\end{eqnarray}
while the KG equations for the auxiliary fields are:
\begin{eqnarray}
  U'' + \left(3 + \xi\right)U' = 6\left(2 + \xi\right) \;, \label{KGDWU}\\
  V'' + \left(3 + \xi\right)V' =-6\left(2 + \xi\right)\Bar{f} \;. \label{KGDWV}
\end{eqnarray}
If we impose initial conditions compatible with a radiation-dominated epoch, where $h_i^2 \sim \Omega_{Ri}$ and $\xi_i \sim -2$, Eqs. \eqref{CosmoDW1} and \eqref{CosmoDW2} provide the two following constraints:
\begin{eqnarray}
        f_i - V_i = -\frac{1}{6}U_i'V_i' - f_i' + V_i' \; , \label{ICCosmoDW1} \\
        -f_i + V_i = -V_i' -f_i'' + \frac{U_i'V_i'}{2} \label{ICCosmoDW2} \; ,
\end{eqnarray}
where in Eq. \eqref{ICCosmoDW2} we used Eq. \eqref{KGDWV} evaluated at $\xi_i = -2$. 
As expected, the value of $U_i$ is unconstrained since it appears on the field equations only through the function $f(U_i)$; to compute the time derivative of the latter we use the chain rule $f' =  \Bar{f}U'$, so that:
\begin{equation} \label{f''}
   f'' = \Bar{\Bar{f}}U'^2 -\Bar{f}U'' \; .
\end{equation}
Evaluating Eq. \eqref{f''} at $N = N_i$ we get:
\begin{equation} \label{f_i''}
    f_i''= \Bar{\Bar{f_i}}{U_i'}^2 - \Bar{f}U_i' \; ,
\end{equation}
where we have used Eq.\eqref{KGDWU} with $\xi_i = -2$.
In order to satisfy the constraint Eqs. \eqref{ICCosmoDW1} and \eqref{ICCosmoDW2} we then make the natural choice $V_i =f(U_i) = U_i'= V_i'= 0$.

In Ref. \cite{Deffayet:2009ca} the authors develop a technique to reconstruct the distortion function starting from any cosmological history; when specialized to the case of the $\Lambda$CDM the best analytical approximation for the distortion function is given by:
\begin{equation} \label{fDW}
    f(U) = 0.245\left[ \tanh{\left(0.350Y + 0.032Y^2 + 0.003Y^3\right) } - 1\right]\; ,
\end{equation}
where $Y= -U + 16.5$. 
Note that the above distortion function satisfies the condition $f(U_i) \simeq 0$ by choosing $U_i = 0$.

\section{Late-times behavior of the models} \label{section3}
We are interested in the late-times behavior of these nonlocal models in order to understand in which cases the final stage of the cosmological history of the universe is similar to a DE-dominated one, and if such a behavior is dependent on the free parameters of the theories.

\subsection*{General scheme}

The general scheme presented here was developed in Ref. \cite{Giani:2019vjf} to study the late-times behavior of the model \cite{Vardanyan:2017kal}. In this section we apply the same ideas to the nonlocal models presented in the previous section. A sketch of the general strategy is the following: we use the fact that the sign of the first derivative of the auxiliary fields is determined by the formal solutions of the KG equations; then,  imposing initial conditions compatible with radiation and matter domination, we are able to understand qualitatively the evolution of the nonlocal fields when matter sources are totally diluted by imposing consistency with the first Friedmann equation. Finally, we insert the asymptotic solution obtained for the fields and their derivatives into the acceleration equation to compute the asymptotic value of $\xi$.  Note that the scheme presented here is only valid considering the initial conditions presented in the previous section and if we make the crucial assumptions $\xi + 2 \geq 0$.\footnote{This is reasonable since we fix the initial conditions during the radiation-dominated era, when $\xi = -2$, and then we want to allow for a matter-dominated epoch, for which $\xi = -3/2$.}

\subsubsection{The qualitative behavior of $U$}
To begin with let us consider the KG equation for the field $U$:
\begin{equation}
    U'' + \left( 3 + \xi \right)U' = 6\left(2+\xi \right) \; ,
\end{equation}
the formal solution for the first derivative of the field is given by:
\begin{equation} \label{FormalU}
    U' = 6e^{-F(N)}\int_{N_i}^{N} d\Bar{N} e^{F(\Bar{N})}\left[2 + \xi\left(\Bar{N}\right) \right] - C_1e^{F(N)} \; ,
\end{equation}
where $C_1$ is an integration constant and where the function $F(N)$ is defined by :
\begin{equation}
    F(N) \equiv \int_{N_i}^{N} d\Bar{N}\left[3+ \xi\left(\Bar{N}\right) \right] \; .
\end{equation}

We choose vanishing initial conditions on $U$ during the radiation-dominated era in order not to spoil the cosmological history; this fixes $C_1 = 0$. 
It is straightforward to realize that since $\xi \geq -2$, then $U'$ is always positive. Moreover, it is easy to prove that it is also limited in the range $0 \leq U' < 6$. Indeed, let us rewrite its solution in the form:
\begin{eqnarray}
    U' &=& 6e^{-F(N)}\int_{N_i}^{N} d\Bar{N} e^{F(\Bar{N})}\left(3 + \xi \right) -6e^{-F(N)}\int_{N_i}^{N} d\Bar{N} e^{F(\Bar{N})}\; \nonumber \\
    &=& 6 -6e^{-F(N)} - 6e^{-F(N)}\int_{N_i}^{N}d\Bar{N}e^{F(N)} \; , 
\end{eqnarray}
from which it is straightforward to realize that $0 \leq U' \leq 6$, since the last two terms on the right hand side of the equation are always negative. Being its first derivative always positive and his initial value vanishing, we can then conclude qualitatively that $U > 0$ always. Note that in the radiation-dominated epoch, $N \sim N_i$, we want $U, U'$ to be vanishing in such a way that it is only when the pressureless matter density is appreciable on cosmological scales that $U$ begins its evolution.

\subsubsection{The Friedmann equation}
Due to the presence of the nonlocal fields the general 0-0 modified Einstein equation can be set in the following form:
\begin{equation}
    \left(1+g(N)_{NL}\right)h^2 = \Omega_{R} + \Omega_{M} +\Omega_{NL} \;,
\end{equation}
where the functions $g(N)$ and $\Omega_{NL}$ are the modifications due to  nonlocal terms. 
Since our main interest is in the late-times behavior of these models, we consider this equation in the regime in which the matter and radiation densities are diluted enough and we can ignore their contribution. The initial conditions set the value $g(N_i)$, and since the signs of the first derivative of the auxiliary fields are determined by the KG equations we are able to estimate the asymptotic form of $g(N)$ trough the cosmological history, and in particular its asymptotic value.

\subsection{Late-times behavior of the $RR$ Model}
To begin with let us consider the $RR$ field equations \eqref{CosmoRR1},\eqref{CosmoRR2} and \eqref{KGRRV} when matter and radiation density are negligible and define $\tilde{V} \equiv 1 -3\gamma V$: 
\begin{eqnarray}
\tilde{V} &=& \frac{\gamma U^2}{4 h^2} + \tilde{V}'\left(\frac{U'}{6} -1 \right) \; , \label{CosmoRR1A}\\
\xi &=& \frac{1}{2\tilde{V}}\left[\frac{3\gamma U}{h^2} - U'\tilde{V}' + 4\tilde{V}'\right] \; , \label{CosmoRR2A}\\
\tilde{V}'' &+& \tilde{V}'\left(3 + \xi\right) = -\frac{3\gamma U}{h^2} \; \label{KGRRVA} .
\end{eqnarray}
The formal solution of Eq. \eqref{KGRRVA} for $\tilde{V}'$ compatible with the initial condition $\tilde{V}'(N_i)=0$ is:
\begin{equation}\label{FormalVRR}
    \tilde V' = -3\gamma e^{-F(N)}\int_{N_i}^{N}d\Bar{N}e^{F(\Bar{N})}\frac{U}{h^2}  \; ,
\end{equation}
From Eq. \eqref{FormalVRR} it is straightforward to realize that $\tilde{V}'$ is always negative since $U$ is always positive, while imposing vanishing initial conditions for the nonlocal fields at early times determines the initial value $\tilde{V} =1 $. On the other hand from the  right hand side of Eq. \eqref{CosmoRR1A} we see that $\tilde{V}$ must be positive and so we can conclude that $ 0\leq \tilde{V} \leq 1 $.\footnote{Note that the parameter $\gamma$ can be considered as positive definite since changing its sign corresponds to switch the sign of the nonlocal interaction term in the Lagrangian. In this case it is more convenient to change the sign of the source term in the equation of the auxiliary field $U$,  in such a way that the product $\gamma U$ is positive definite. Here we are neglecting the radiation and matter contribute which are anyway also positive definite.} This last argument tells us then that $\tilde{V}'$ must also vanish at late-times, or it would push $\tilde{V}$ to negative values. On the basis of these considerations we can conclude that at late-times we have
\begin{equation} \label{RRasyVV'}
    \tilde{V} \sim \frac{\gamma U^2}{4 h^2} \; , \qquad    \tilde{V}' \sim 0 \; ,
\end{equation}
and using the above results in Eq. \eqref{CosmoRR2A} we get:
\begin{equation} \label{AxiRR}
    \xi \sim \frac{3\gamma U}{4h^2} \frac{4h^2}{\gamma U^2} \sim \frac{1}{U} \rightarrow 0 \; ,
\end{equation}
where the last limit holds true since $U$ diverges.\footnote{Note that $U$ cannot reach a constant value since $U' = 0$ is possible only for $\xi = -2$, and we easily see from Eq. \eqref{AxiRR} that at late-times $\xi > 0$. }
We have then shown analytically that the effective equation of state of the $RR$ model approaches asymptotically  $\omega_{\rm eff} \rightarrow -1$.

\subsection{Late-times behavior of the $m^2\frac{1}{\Box}R$ model} 
This subsection closely follows the treatment made by the authors in Ref. \cite{Giani:2019vjf}.
When matter is diluted Eqs. \eqref{CosmoVAAS1} and \eqref{CosmoVAAS2} become:
\begin{eqnarray}
    	3\tilde{V} &=& \frac{\gamma U}{2h^2} - \frac{V'}{2}(6 - U') \; ,\label{LTCosmoVAAS1}\\
        \xi &=& -3 + \frac{1}{\tilde V}\left[-\tilde{V}' + \frac{\gamma(1 + U)}{2h^2}\right]\;,
\end{eqnarray}
while the KG equation for $\tilde{V}$ \eqref{KGVAASV} is given by:
\begin{equation}
    \tilde{V}'' + \tilde{V}\left(3 + \xi \right)\tilde{V}' = -\frac{\gamma}{h^2} ,
\end{equation}
and its formal solution is:
\begin{equation}\label{FormalVVAAS}
    \tilde V' = -e^{-F(N)}\int_{N_i}^{N}d\Bar{N}e^{F(\Bar{N})}\frac{\gamma}{h^2}  \; .
\end{equation}
Since in this model $U > 0$, $ 0 < U' < 6$, while $\tilde{V}' < 0$ from Eq. \eqref{FormalVVAAS}, we can conclude from \eqref{LTCosmoVAAS1} that $\tilde V > 0$. On the other hand, $\tilde{V}' < 0$ tells us that $\tilde V$ always decreases. So, in order for $\tilde V$ to decrease from one to zero, without becoming negative, we need that at late-times $m^2/h^2 \ll 1$ and $\tilde{V}' \sim 0$. On the basis of this argument, we can conclude that:
\begin{equation}
	3\tilde V \sim \frac{\tilde{m}^2U}{2h^2}\;,
\end{equation}
and finally:
\begin{equation}
	\xi \sim -3 + \frac{1}{\tilde V}\frac{\tilde{m}^2U}{2h^2} \sim 0\;,
\end{equation}
and we have shown that asymptotically $\omega_{\rm eff} \rightarrow -1$.

\subsection{Late-times behavior of the $\frac{1}{\Box^2}R$ model}
In this case defining $\tilde{Z}= 1-\frac{3}{2}\gamma Z$ we can rewrite the late-times Friedmann equations \eqref{CosmoB1} \eqref{Cosmob2}, when matter is completely diluted, as:
\begin{eqnarray}
        \tilde{Z} &=& \frac{\gamma}{4h^2}\left(UW + V \right) -\tilde{Z}'\left(1 - \frac{U'}{6}\right) - \frac{\gamma W' V'}{4} \; , \label{CosmoB1A}\\
        \xi &=& \frac{1}{2\tilde{Z}}\left[\frac{3\gamma W}{2h^2} + 4\tilde{Z}' -\tilde{Z}'U' + \frac{3\gamma V'W'}{2} \right] \; \label{CosmoB2A} ,
\end{eqnarray}
while the KG equation \eqref{KGBZ} for $\tilde{Z}$ is:
\begin{equation}
    \tilde{Z}'' + \left(3+\xi \right)\tilde{Z}' = -\frac{3\gamma W}{2h^2} \; ,
\end{equation}
whose formal solution for $\tilde{Z}'$ is given by:
\begin{equation} \label{FormalZB}
 \tilde Z' = -3\gamma e^{-F(N)}\int_{N_i}^{N}d\Bar{N}e^{F(\Bar{N})}\frac{W}{2h^2}  \; .
\end{equation}
We write for convenience also the formal solutions for $V'$ and $W'$:
\begin{eqnarray}
          V' = e^{-F(N)}\int_{N_i}^{N}d\Bar{N}e^{F(\Bar{N})}\frac{U}{h^2}  \; , \label{FormalVB}\\
          W' = e^{-F(N)}\int_{N_i}^{N}d\Bar{N}e^{F(\Bar{N})}\frac{1}{h^2}  \; \label{FormalWB} .
\end{eqnarray}
Since for our choice of initial conditions $W_i = 0$ and since $W' > 0$ we can conclude that $W > 0$. This implies that $\tilde{Z}'< 0$; on the other hand Eqs. \eqref{FormalVB} and \eqref{FormalU} imply $V'>0$, $0< U' <6$ and $U > 0$. 

In order to understand the behavior of $\tilde{Z}$ let us define the function $X$:
\begin{equation}
   X \equiv \frac{UW + V}{h^2} - W'V' \; ,
\end{equation}
 taking its time derivative and using Eqs. \eqref{KGBW} and \eqref{KGBV} we are able to set up a differential equation for $X$:
 \begin{equation}\label{diffeqx}
   X' + 2\xi X = \frac{U'W}{h^2} + 6V'W' \; .
 \end{equation}
 The formal solution of Eq. \eqref{diffeqx} is given by:
\begin{equation} \label{xsol}
    X\left(N\right) = \frac{1}{h^2(N)}\int_{N_i}^{N} d\Bar{N} \left(U'W + 6h^2V'W'\right) - C_X h^2(N) \; ,
\end{equation}
where $C_X$ is an integration constant.
 Since $X(N_i)= 0$ we can conclude from Eq. \eqref{xsol} that $X(N) > 0$.
 This in turns implies that the right hand side of Eq. \eqref{CosmoB1A} it's always positive, and we can conclude that, asymptotically, $\tilde Z > 0 $.

Since $\tilde{Z}$ is positive definite, we must have asymptotically $\tilde{Z}' \rightarrow 0$. It is straightforward to realize from Eq. \eqref{FormalZB} that this is possible only if $h^2$ is a monotonic growing function that grows faster than $W/2$. On the other hand, $W$ is also a monotonic growing function since $W' > 0$. In particular, since $h^2$ grows faster than $W$, it also grows faster then a constant, and so we conclude from Eq. \eqref{FormalWB} that $W' \rightarrow 0$.
Using the latter in Eq. \eqref{CosmoB1A} we are left with: 
\begin{equation}
    \tilde{Z} \sim \frac{\gamma}{4h^2}\left(V + WU \right) \; .
\end{equation}
Using the above result in Eq. \eqref{CosmoB2A} we finally obtain:
\begin{equation}
    \xi \sim \frac{3}{\frac{V}{W} + U} \sim 0 \; ,
\end{equation}
then once again we have $\omega_{\rm eff} \rightarrow -1$.

\subsection{The late-times behavior of  the DW model }
In order to study qualitatively the dynamic of the DW model at late-times we have first of all to understand qualitatively the behavior of the free function $f(U)$ defined in Eq.\eqref{fDW}, since it enters directly in the Friedmann equations and also rules the dynamics of the localized field $V$. It is straightforward to realize from Eq. \eqref{fDW} that $(-2)(0.245) < f < 0$ and $\Bar{f} < 0$ , and that \ $\Bar{f} \rightarrow 0$ when $U \rightarrow \infty$.

The formal solution of \eqref{KGDWV} for $V'$ is:
\begin{equation} \label{FormalVDW}
    V' =  -6e^{-F(N)}\int_{N_i}^{N}d\Bar{N}e^{F(\Bar{N})}\left(2+\xi\right)\Bar{f}  \; .
\end{equation}
Since $\Bar{f} < 0$ and $\xi > -2$ from Eq. \eqref{FormalVDW} it is straightforward to realize that we have at all times $V'> 0$. Since we impose initial conditions in such a way that during the radiation-dominated epoch $V$ is vanishing, we also can conclude that $V > 0$. At late-times, when matter is completely diluted Friedmann equations \eqref{CosmoDW1} and \eqref{CosmoDW2} become:
\begin{eqnarray}
        V &=& -V'\left(1-\frac{U'}{6}\right)  +\Bar{f}U' + f + 1 \; , \label{CosmoDW1A} \\
        \left(2\xi + 3\right)\left(1+f-V\right) &=& V'' - f'' + \left(V' - f'\right)\left(2 + \xi  \right) + \frac{U'V'}{2}\; \label{CosmoDW2A} .
\end{eqnarray}
Note that the first two terms in the right hand side of Eq.  \eqref{CosmoDW1A} are strictly negative since, $V' > 0$ and $\Bar{f}<0$, while $f+1 > 0$. On the other hand $V' >0$ implies that $V$ is a monotonic function, and we are left with two cases; either $U$ diverges, in which case $\Bar{f} \rightarrow 0$ and $f \rightarrow (-2)(0.245)$, or $U \rightarrow const$, in which case $U' \rightarrow 0$, $f'\rightarrow 0$ and $f + 1 \rightarrow const$. In both cases consistency requires that  $V' \rightarrow 0$, or  $V$ will be a decreasing function and so $V' < 0$. Thus we can conclude that asymptotically:
\begin{equation}
    V \sim \Bar{f}U' +  f + 1 \; .
\end{equation}
Using the above in \eqref{CosmoDW2A} we obtain finally:
\begin{equation}
    \xi \sim \frac{U' - 12 -\frac{f''}{\Bar{f}}}{6-U'}  \; ,
\end{equation}
which is in general non-vanishing.
Note also that:
\begin{equation}
    \frac{f''}{\Bar{f}} = \frac{(\Bar{f}U')'}{\Bar{f}}= U'' + \frac{\Bar{\Bar{f}}}{\Bar{f}}U'^2 \; ;
\end{equation}
 and we can conclude that if $U \rightarrow \infty$ the term $\Bar{\Bar{f}}/\Bar{f} \rightarrow - \infty$, while $U''$ cannot diverge since $0<U'<6$, so in this case the asymptotic effective equation of state $\omega_{\rm eff} \rightarrow \infty$. On the other hand, if $U \rightarrow const$, we have $U' = U'' \rightarrow 0$ and we are left with $\xi \rightarrow -2$, in such a way that the effective equation of state approaches one of radiation type.
 
We have then shown that in the DW model at late-times $\omega_{\rm eff} \neq -1$.

\section{Discussion and Conclusions} \label{section4}
In this work we study the asymptotic behavior of some nonlocal modifications of gravity that involve functions of the inverse D'Alembertian operator acting on the Ricci scalar. In particular, we show that in the models in which the term $U = (1/\Box) R$ appears explicitly in the field equations, if we impose vanishing initial conditions for the auxiliary fields in order to be in a natural way compatible with the radiation-dominated epoch,  the effective asymptotic equation of state always tend to $-1$. We want to stress that our results are analytic, and the method used is the same independently from the model.

On the other hand we find that for the DW model, where he field $U$ does not appear explicitly in the field equation, the asymptotic late-times behavior is different. In this model the field $U$ appears in the equation of motion only inside the argument of a hyperbolic tangent; this means that even if its dynamics pushes $U$ to $\rightarrow \infty$ like in the other models, this divergence do not appear in the field equations since $-1\leq \tanh{U} \leq 1$. Note that the field $U$ is the only one among the auxiliary fields that were introduced that is allowed to diverge, indeed all the source terms of the other KG equations vanish as $\sim h^{-2}$, with the exception of Eqs. \eqref{KGBV},\eqref{KGDWV} whose source terms, however, cannot diverge since the former must vanish asymptotically in order to be compatible with Eq. \eqref{CosmoB1A} and the latter vanishes as $-U^2\cosh^{-2}\left(-U^3\right)$. To summarize, the structure of the source terms of the KG equations implies that only the auxiliary field related to the non local term $1/\Box R$ is still dynamical asymptotically, and diverges, while the auxiliary fields related to the Lagrange multipliers used to localize the theories freeze  and approach a constant value.   

It is important to remark that the above conclusions strongly depend on the choice of initial conditions. Indeed, our method relies on the observation that by using Friedmann equations we can constrain the sign of the auxiliary fields, while their KG equations provide constraint on the sign of the first derivatives for our choice of initial conditions. 
As an example, let us consider the $RR$ model; if we choose a negative initial value in Eq. \eqref{KGRRV} for the field $U$  then $V' >0 $ and Eq. \eqref{RRasyVV'} does not hold anymore. This situation correspond to the evolution Path B described in Ref. \cite{Nersisyan:2016hjh}, for which at late-times $\omega_{\rm eff}\rightarrow 1/3$. However, our  analysis still holds for any choice of initial conditions with non-vanishing but positive values of $U$. As discussed in Ref. \cite{Belgacem:2017cqo} there are fundamental motivations that justify processes during the inflationary epoch that result in a huge non-vanishing positive values for the field $U$ in the RD epoch.

We want to stress that an asymptotic behavior of the form $\omega_{\rm eff} \rightarrow -1 $ is a remarkably feature for a model that wants to be competitive with the $\Lambda$CDM. Indeed in such  models, and in the $\Lambda$CDM, the so called \textit{Coincidence Problem} \cite{Velten:2014nra} is less severe (if not a problem at all, depending from the personal perspective), since at some point of its history independently from the initial conditions the universe always passes trough a phase in which the matter and DE densities are of the same order and then DE starts to dominate, which in the standard model in terms of cosmic time accounts for at least the last 3.5 billions of years. On the other hand in the nonlocal models considered here we have to deal with a different sort of coincidence. Indeed the Hubble function reaches a minimum when the nonlocal fields cosmological density starts to dominate, and the occurrence of this is roughly today. This occurrence looks to us  coincidental at least as the one present in $\Lambda$CDM. 

\subsection*{Acknowledgements}
The authors are grateful to Luca Amendola for useful comments and suggestions.
This study was financed in part by the \emph{Coordena\c{c}\~ao de Aperfei\c{c}oamento de Pessoal de N\'ivel Superior} - Brazil (CAPES) - Finance Code 001. OFP thanks the Alexander von Humboldt foundation for funding and both the authors are grateful to the Institute for Theoretical Physics of the Heidelberg University for kind hospitality. 

\bibliographystyle{unsrturl}
\bibliography{LT-NL}
\end{document}